\documentclass[10pt]{iopart}
\usepackage{psfig}  
\begin{document}\hbadness=10000
\title[Chemical Nonequilibrium]{Chemical Nonequilibrium in\\ High
Energy Nuclear Collisions}
\author{Jean Letessier}
\address{LPTHE, Universit\'e Paris 7, 2 pl.\,Jussieu, F--75251 Cedex 05
}
\author{Johann Rafelski}
\address{Department of Physics, University of Arizona, Tucson, AZ 85721
}
\begin{abstract}
Strange particles produced in S--Au/W/Pb 200 A GeV and Pb--Pb 158 A GeV 
reactions are described invoking final hadronic phase space in thermal 
equilibrium, but allowing chemical non-equilibrium. Several sets of statistical 
freeze-out parameters are obtained for each system, invoking different models 
of dense matter. We show that only when allowing for strange and
non-strange flavor abundance non-equilibrium, a statistically 
significant description of the experimental results is obtained.
Physical properties of the fireball at chemical freeze-out 
condition are evaluated and considerable universality 
of hadron freeze-out between the two different collision systems 
is established. The relevance of the Coulomb effect in the highly
charged Pb--Pb fireballs for the chemical analysis are discussed. 
The influence of explosive collective matter flow is also described. 
\end{abstract}
%

%
\section{Chemical non-equilibrium statistical model}\label{secnoneq}
\subsection{Introductory remarks}
We report here on the recent progress we made with the 
thermal model freeze-out analysis of the CERN-SPS 200 A GeV 
Sulphur beam reactions with laboratory stationary 
`heavy' targets, such as Gold, Tungsten or Lead nuclei \cite{LR98}, 
and we also present the current status of our ongoing effort 
to understand the results from CERN 158 A GeV Lead beam  
Pb--Pb collisions \cite{fit97,LRPb98}. These reactions occur at an
energy  $E_{\rm CM}=\sqrt{s}/B=8$--$9\,{\rm GeV}\simeq9m_{\rm N}c^2$
per participating  nucleon in the center of momentum frame. 
This high energy available 
materializes in form of high hadronic particle multiplicity which 
we are aiming to interpret. In this work we assume local 
thermal ({\it i.e.,} energy equipartition) equilibrium 
\cite{HAG,HAGBB,acta96,BSWX96} reached in  a relatively small and dense 
volume of highly excited hadronic matter, the `fireball'.
One can argue that the accessibility
of many degrees of freedom, as expressed by high specific entropy
content, validates thermal equilibrium approach. As of now 
there is no established theoretical argument for the rapid kinetic
equilibration process in highly excited, dense hadronic matter. 
However, this seems to be the case: consider that as seen  in the
results of the precise measurements made by 
experiments WA85/WA94/WA97 \cite{Eva96,WA97}, the transverse mass 
spectra are nearly identical for particle-antiparticle pairs 
where particles comprise some quarks brought into the reaction, 
{\it e.g.},  $\overline\Lambda$--$\Lambda$, $\Xi$--$\overline\Xi$. When 
modeling  the production of these particles in microscopic models,
we encounter in general vastly different $m_\bot$ spectra \cite{Bas98}.

The fireball undergoes complex {\it chemical} evolution 
until at some stage final state particle abundances freeze-out. 
We refer here to the stage in evolution of the
fireball at which density has dropped to the level that
in subsequent collisions particle abundances remain unchanged.
The mechanisms of chemical equilibration in which particle numbers change
are today theoretically better understood than are mechanisms responsible for
what is believed to be much faster to establish thermal (kinetic) equilibration, 
where momentum exchange between existent particles is the key mechanism.
Recall that it has been the allowance of non-equilibrium 
chemical abundance for strange quarks which permitted to analyze accurately 
the experimental strange particle abundance data and to characterize
the properties of the particle source \cite{acta96,Raf91,Let95,Sol97,BGS98}. 
Here our primary result, concluded from the success of the description of the 
experimental data, is that chemical non-equilibrium both for light
and strange quarks is a necessary requisite 
for the understanding of the experimental particle abundances. 
The corresponding technical refinement not present in
earlier work  is that we introduce a parameter, the light quark
phase space occupancy (see below) $\gamma_{\rm q}$ 
to describe the light quark chemical non-equilibrium\cite{LR98,LRflow98}. 
We did not  previously consider simultaneously a   interpretation of 
strange and non-strange particles and hence the need to allow for 
light quark chemical non-equilibrium was not visible to us: the strange 
quark phase space occupancy $\gamma_{\rm s}$ was determined {\it relative} 
to $\gamma_{\rm q}$, which we now understand is also off-equilibrium. Since 
we now can accurately describe abundances of strange as well as non-strange
hadrons, we combine in the present analysis the strangeness diagnostic tools
of dense matter with the entropy enhancement \cite{Let95,Let93}.
We side-step here initially, mainly to keep the number of parameters to
a minimum, the need to study collective matter flows originating in both, 
the  memory of the initial `longitudinal' collision momentum, and 
the explosive disintegration driven by the 
internal pressure of compressed hadronic matter. However, we will present in
section~\ref{flowsec} a short account of the results obtained with
such effects, and illustrate changes that arise in our present study. 

\subsection{Parameters and their physical meaning}
We employ the local thermal  equilibrium method, and thus use a local 
freeze-out temperature $T_{\rm f}$. Regarding chemical equilibration, we  
will recognize two different types, relative and absolute.
We speak of relative chemical equilibration occurring 
through quark exchange reactions described in terms of 
fugacities $\lambda_i,\,i=u,\,d,\,s$. In general we will
not distinguish between the two light $ q=u,\,d$ quarks, and
thus we use two fugacities only. The 
hadronic particle fugacities are obtained  from the valance 
quark fugacities. However, the fugacities 
$\lambda_i$ do not regulate the total number of 
quark-pairs present, and this number 
 has to be controlled  by a new parameter, the phase space occupancies
 $\gamma_i,\,i=u,\,d,\,s$ --- again, we shall not distinguish between
the two light flavors. 

To understand the role of $\gamma_i$ note that any 
compound particle comprising a
particle-antiparticle pair is not controlled in abundance by a fugacity,
since the formation of such particles does not impact the
conservation laws. The abundance of, {\it e.g.,} neutral
pions comprises normally no (quark) fugacity at all. 
This abundance is thought to be regulated solely
by temperature. This of course implies the tacit assumption of 
absolute chemical equilibrium. However, the effective fugacity of
quarks is  $\lambda_i\gamma_i$ and antiquarks $\lambda_i^{-1}\gamma_i$,
and thus with the introduction of $\gamma_i$ we can  control
pair abundance independently of other  properties of the system. 
The proper  statistical physics foundation of  $\gamma_i$ 
is obtained considering the maximum entropy principle 
for evolution of physical systems. In such a study it has
been determined that while the equilibrium 
limit $\gamma_i\to1$ maximizes the specific
chemical entropy, this maximum is very shallow \cite{entro}, indicating that 
a system with dynamically evolving physical properties such as 
the occupied volume will in general find more 
effective paths to increase entropy, than offered by the establishment of 
the absolute chemical equilibrium. 

The dynamical theory for $\gamma_{\rm s}$ has been one of the 
early cornerstones of the proposal to use strangeness as signature of
deconfinement \cite{RM82,BZ82}. 
A time dependent build up of chemical abundance was first 
considered in the context of microscopic strangeness production in 
QGP, after it was realized that 
strange flavor production occurs at
the same time scale as the collision process. More generally,
one must expect, considering the time scales, that all quark flavors
will not be able to exactly follow  the rapid evolution in
time of dense hadronic matter. Moreover, fragmentation of gluons
in hadronizing QGP can contribute additional quark pair
abundance, conveniently described by the factor $\gamma_i$. It is
thus to be expected that also  for light quarks
the chemical phase space occupancy factor $\gamma_q\ne 1$.
Introduction of the factor $\gamma_q$ leads to
a  precise chemical description of the S--Au/W/Pb 200 A GeV
collisions \cite{LR98}, which was not possible before. 
The tacit choice $\gamma_q= 1$ has  not  allowed previously
to distinguish the different reaction scenarios in Pb--Pb
collisions leading to contradictory results \cite{fit97,BGS98}.
Introduction of $\gamma_q$, along with
improvement in experimental precision, and a greater data sample, 
allows to develop the here presented precise analysis.

\subsection{Hadronic phase space}
The evaluation of the final particle yields follows the pattern 
established in our earlier work (see, {\it e.g.}, 
\cite{acta96}).   The relative number of primary 
particles freezing out from a source is obtained 
noting that the  fugacity and phase space occupancy 
 of a composite hadronic  particle is  expressed  
by its constituents and that the probability to find all 
$j$-components contained within  the $i$-th  emitted particle is:
\begin{equation}\label{abund}
N_i\propto \prod_{j\in i}\gamma_j\lambda_je^{-E_j/T}\,,
\qquad\lambda_i=\prod_{j\in i}\lambda_j\,,
\qquad \gamma_i=\prod_{j\in i}\gamma_j\,.
\end{equation}
The unstable hadronic resonances 
 are allowed to disintegrate and feed the stable hadron spectra.  
Full phase  space coverage or central rapidity region $|y-y_{\rm CM}|<0.5$, 
is considered, where as usual the energy of the particles is expressed by
 $$E_j=\sqrt{m_j^2+p^2}=\sqrt{m_j^2+p_\bot^2}\cosh y\,, $$ 
and $y_{\rm CM}$ is the center of momentum rapidity of the colliding nuclei.

Hadron spectra are not significantly 
deformed by cascading decays of heavier resonances at sufficiently high
$p_\bot$, and all small acceptance particle ratios we consider are
thus chosen to satisfy well this criterion. Therefore, in our approach,
we first evaluate the partial multiplicities
for different hadrons within the acceptance domain,  and allow these to 
decay to obtain the contributing fraction. In principle the 
approach must be to take the momentum distribution of primary particles, have 
the momentum distribution cascade into secondaries which may again cascade.
Only the  final abundance  is then kinematically cut to the experimental
acceptance. Unless a random simulation is carried out (unsuitable for rarely 
produced particles we are interested in), this approach is 
completely impossible numerically: at each level of cascading there is a
new level of averaging over the phase space in addition to the one
needed  when treating the flow, see below, and thus normally 
only ONE decay step is accounted for. Our approach thus has the advantage
of incorporating multi-step cascading effects, at the cost of somewhat 
inaccurate implementation of cuts. However, we believe, and have checked in so
far possible,  that the theoretical uncertainties our approach introduces are
much smaller than the experimental errors. 

Once the parameters  $T_{\rm f},\,\lambda_{\rm q},\,\lambda_{\rm s},\,
\gamma_{\rm q},\,\gamma_{\rm s}$ are determined from the   
particle yields available, we can evaluate the properties of the 
entire hadronic particle phase space and obtain the 
physical properties of the system, such as, {\it e.g.}, 
energy and entropy per baryon, strangeness content. 
Even though we are describing a free streaming gas of emitted particles, we 
can proceed as if we were evaluating partition function of system with the 
phase space distribution described 
by the statistical parameters, given that just in an earlier instant
in a gedanken experiment we have a cohesive,
interacting system. We have implemented all relevant hadronic states and 
resonances in this approach and have also included quantum statistical 
corrections, allowing for first
Bose and Fermi distribution corrections in the hadron abundances and in
the phase space content. These corrections influence favorably the quality 
of the agreement between theory and experiment.

\subsection{Inclusion of collective flow}\label{subsecflow}
Most of our analysis will be carried out under the assumption that the
effects of matter flow can be accommodated later, since the $4\pi$ particle 
ratios are not  influenced, and the ratios of compatible particles
({\it i.e.}, those dragged along by matter in the same manner, {\it viz}, strange
baryons and antibaryons) are not affected significantly. While this is
correct in general, the precision of the experimental data at this time 
requires a more detailed study, for the flow also depends on particle
mass and strange (anti)baryons differ by nearly 50\% comparing $\Lambda$
with $\Omega$. Thus, in section~\ref{flowsec},  we shall briefly discuss 
the influence on our analysis  of the collective matter 
flow flow velocity $\vec v_{\rm c}$ at chemical freeze-out.  
Several different schemes to implement flow 
were studied previously \cite{Hei92}.  As our example we adopt here a radial 
expansion model and  consider the causally disconnected domains 
of the dense matter fireball  to be synchronized at the 
instance of collision --- in other words the time of freeze-out 
is for all volume elements a common constant time in the CM frame. 
The freeze-out occurs at the surface of 
the fireball simultaneously in the CM frame, but not necessarily 
within a short instant of CM-time. Since we do not use
a radial profile of the flow velocity in the fireball, which
amounts to the tacit assumption that all hadrons are born at 
same surface velocity. This would be the case if the 
chemical freeze-out condition were also the QGP-hadronization
condition.

Within this approach  the spectra and thus also multiplicities 
of particles emitted are obtained replacing 
the Boltzmann factor in Eq.\,(\ref{abund}) by:
\begin{equation}\label{abundflow}
e^{-E_j/T}\to \frac1{2\pi}\int d\Omega_v
  \gamma_v(1+\vec v_{\rm c}\cdot \vec p_j/E_j)
  e^{-{{\gamma_vE_j}\over T}
    \left(1+\vec v_{\rm c}\cdot \vec p_j/E_j\right)} \,,
\end{equation}
where as usual $\gamma_v=1/{\sqrt{1-\vec v_{\rm c}^{\,2}}}$\,. 
Eq.\,(\ref{abundflow}) can  be intuitively obtained by a Lorentz
transformation between an observer on the surface of
the fireball, and one at rest in laboratory frame. In 
certain details the results we obtain
confirm the applicability  of this simple approach.  

As can be seen in Eq.\,(\ref{abundflow}) we need to carry 
out an additional two dimensional  (half-sphere)  surface 
integral with the coordinate system fixed by an arbitrary, but fixed 
(collision) axis which defines the transverse particle momentum.
Just a one dimensional numerical integration over one of the surface 
angles needs to be carried out. To obtain the particle spectra as function 
of $m_\bot$ we need also to integrate over rapidity $y$\,. This rapidity
integration can be approximated for  a narrow rapidity interval 
using  the error function.

\section{Experimental results and the data analysis}
We consider for S--Au/W/Pb reactions 
18 data points listed in table~\ref{resultsw} 
(of which three comprise  the $\Omega$'s). For Pb--Pb we address 
here 15 presently available particle yield ratios 
listed in table~\ref{resultpb}
(of which four  comprise  the $\Omega$'s). 
We believe to have included in our discussion most if not all 
particle multiplicity results available presently.
\begin{table}[tb]
\caption{\label{resultsw}
Particle ratios studied in our analysis for  S--W/Pb/Au reactions: 
experimental results with  references and kinematic cuts are given, 
followed by columns showing results for the different strategies  of analysis B--F.
Asterisk~$^*$ means a predicted result (corresponding data is not 
fitted). The experimental results here considered are 
from\\ \footnotesize
$^1$ {S.\,Abatzis {\it et al.}, WA85 Collaboration, {\it Heavy Ion Physics} {\bf 4}, 79 (1996).} \\
$^2$ {S.\,Abatzis {\it et al.}, WA85 Collaboration, {\it Phys.\,Lett.}\,B {\bf 347}, 158 (1995).} \\ 
$^3$ {S.\,Abatzis {\it et al.}, WA85 Collaboration, {\it Phys.\,Lett.}\,B {\bf 376}, 251 (1996).} \\ 
$^4$ {I.G.\,Bearden {\it et al.}, NA44 Collaboration, {\it Phys.\,Rev.}\,C {\bf 57}, 837 (1998).}  \\
$^5$ {D.\,R\"ohrich for the NA35 Collaboration, {\it Heavy Ion Physics} {\bf 4}, 71 (1996).} \\ 
$^6$ S--Ag value adopted here: {T.\,Alber {\it et al.}, NA35 Collaboration, \\ \hspace*{0.15cm}
{\it Eur.\,Phys.\,J.} C {\bf 2}, 643 (1998); [hep-ex/9711001].}\\
$^7$ A. Iyono {\it et al.}, EMU05 Collaboration, {\it Nucl.\,Phys.\,}A {\bf 544},
455c (1992) and  \\ \hspace*{0.15cm}
 Y. Takahashi {\it et al.}, EMU05 Collaboration,  private communication.}
\small\begin{center}
\begin{tabular}{|lclc|lllll|}
\hline\hline    
 Ratios & $\!\!\!\!$Ref. & Cuts[GeV] & Exp.Data      & B  &C  &D & D$_s$ & F  \\
\hline
${\Xi}/{\Lambda}$                           & 1 &   
$1.2<p_{\bot}<3$   &0.097 $\pm$ 0.006                & 0.16 & 0.11 & 0.099 &0.11 & 0.10 \\
${\overline{\Xi}}/{\bar\Lambda}$            & 1 &  
$1.2<p_{\bot}<3$ &0.23 $\pm$ 0.02                    & 0.38  & 0.23  & 0.22   &0.18 & 0.22 \\
${\bar\Lambda}/{\Lambda}$                   & 1 &
$1.2<p_{\bot}<3$ &0.196 $\pm$ 0.011                  & 0.20 & 0.20 & 0.203 &0.20 & 0.20 \\
${\overline{\Xi}}/{\Xi}$                    & 1 &    
$1.2<p_{\bot}<3$    &0.47 $\pm$ 0.06                 & 0.48  & 0.44  & 0.45  &0.33 & 0.44  \\
${\overline{\Omega}}/{\Omega}$              & 2 &
$p_{\bot}>1.6$&0.57 $\pm$ 0.41                       &1.18$^{*}$&0.96$^{*}$& 1.01$^{*}$&0.55$^{*}$ & 0.98 \\
$\Omega+\overline{\Omega}\over\Xi+\bar{\Xi}$& 2 &
{$p_{\bot}>1.6$}&0.80 $\pm$ 0.40                     &0.27$^{*}$&0.17$^{*}$& 0.16$^{*}$&0.16$^{*}$& 0.16  \\
${K^+}/{K^-}$                               & 1 & 
{$p_{\bot}>0.9$}         &1.67  $\pm$ 0.15           & 2.06  & 1.78  &  1.82 &1.43 & 1.80  \\
${K^0_{\rm s}}/\Lambda$                     & 3 & 
{$p_{\bot}>1$}      &1.43  $\pm$ 0.10                & 1.56  & 1.64  &  1.41 &1.25 & 1.41  \\
${K^0_{\rm s}}/\bar{\Lambda}$               & 3 &
{$p_{\bot}>1$}  &6.45  $\pm$ 0.61                    & 7.79  & 8.02  &  6.96 &6.18 & 6.96 \\
${K^0_{\rm s}}/\Lambda$                     & 1 & 
{$m_{\bot}>1.9$}    &0.22  $\pm$ 0.02                & 0.26  & 0.28  &  0.24 &0.24 & 0.24 \\
${K^0_{\rm s}}/\bar{\Lambda}$               & 1 &
{$m_{\bot}>1.9$}  &0.87  $\pm$ 0.09                  &1.30   & 1.38  &  1.15 &1.20 & 1.16 \\
${\Xi}/{\Lambda}$                           & 1 &
{$m_{\bot}>1.9$}       &0.17 $\pm$ 0.01              & 0.27  & 0.18  &  0.17 &0.18 & 0.17  \\
${\overline{\Xi}}/{\bar\Lambda}$            & 1 &
{$m_{\bot}>1.9$}&0.38 $\pm$ 0.04                     & 0.64  & 0.38  &  0.38 &0.30 & 0.37 \\
$\Omega+\overline{\Omega}\over\Xi+\bar{\Xi}$& 1 &
{$m_{\bot}>2.3$}&1.7 $\pm$ 0.9                       &0.98$^{*}$&0.59$^{*}$ & 0.58$^{*}$&0.52$^{*}$& 0.58  \\
$p/{\bar p}$                                & 4 &
Mid-rapidity &11 $\pm$ 2\ \                          & 11.2  & 10.1  & 10.6  &7.96 & 10.5 \\
${\bar\Lambda}/{\bar p}$                    & 5 & 
4 $\pi$   &1.2 $\pm$ 0.3                             & 2.50  & 1.47  & 1.44  &1.15 & 1.43  \\
$h^-\over p-\bar p$                         & 6 &        
4 $\pi$   &4.3 $\pm$ 0.3                             & 4.4   & 4.2  & 4.1    & 3.6 & 4.1\\
$h^+-h^-\over h^++h^-$                      & 7 &  
4 $\pi$   &0.124 $\pm$ 0.014                         & 0.11 & 0.10 & 0.103 &0.09 & 0.10 \\
\hline
 & & $\chi^2_{\rm T}$ & & 264 & 30 & 6.5 & 38 & 12\\
\hline\hline
\end{tabular}
\end{center}
 \end{table}
\begin{table}[tb]
\caption{\label{resultpb}
Particle ratios studied in our analysis for  Pb--Pb reactions: 
experimental results with  references and kinematic cuts are given, 
followed by columns showing results for the different strategies of analysis B--F.
Asterisk~$^*$ means a predicted result (corresponding data is not 
fitted or not available). The experimental results here considered are 
from\\ \footnotesize
$^1$ {I.\,Kr\'alik, for the WA97 Collaboration, {\it Nucl. Phys.} A (1998)\\ \hspace*{0.15cm}
(presented at Tsukuba QM1998 meeting).}\\
$^2$ {G.J.\,Odyniec, for the NA49 Collaboration, {\it J. Phys.} G {\bf 23}, 1827 (1997).}\\
$^3$ {P.G.\,Jones, for the NA49 Collaboration, {\it Nucl. Phys.} A {\bf 610}, 188c (1996).}\\
$^4$ {F.\,P\"uhlhofer, for the NA49 Collaboration, {\it Nucl. Phys.} A (1998) \\ \hspace*{0.15cm}
(presented at Tsukuba QM1998 meeting).}\\
$^5$ {C.\,Bormann, for the NA49 Collaboration, {\it J. Phys.} G {\bf 23}, 1817 (1997).}\\
$^6$ {G.J.\,Odyniec,  {\it Nucl. Phys.} A  (1998) 
(presented at Tsukuba QM1998 meeting).}\\
$^7$ {D. R\"ohrig,  for the NA49 Collaboration, \\ \hspace*{0.15cm}
``Recent results from NA49 experiment on 
Pb--Pb collisions at 158 A GeV'',\\ \hspace*{0.15cm}
see  Fig. 4, in proc. of EPS-HEP Conference, Jerusalem, Aug. 19-26, 1997.}\\
$^8$ {A.K.\,Holme, for the WA97 Collaboration, {\it J. Phys.} G {\bf 23}, 1851 (1997).}
}\small\begin{center}
\begin{tabular}{|lclc|lllll|}
\hline\hline
 Ratios & $\!\!\!\!$Ref. & Cuts[GeV] & Exp.Data      & B  &C  &D & D$_s$ & F \\
\hline
${\Xi}/{\Lambda}$                               &\footnotesize{1} &$p_{\bot}>0.7$&0.099 $\pm$ 0.008
&  0.138  &  0.093  &  0.095  &  0.098 &  0.107  \\
${\overline{\Xi}}/{\bar\Lambda}$                &\footnotesize{1} &$p_{\bot}>0.7$&0.203 $\pm$ 0.024
&  0.322  &  0.198  &  0.206  &  0.215 &  0.216  \\
${\bar\Lambda}/{\Lambda}$                       &\footnotesize{1} &$p_{\bot}>0.7$&0.124 $\pm$ 0.013
&  0.100  &  0.121  &  0.120  &  0.119 &  0.121  \\
${\overline{\Xi}}/{\Xi}$                        &\footnotesize{1} &$p_{\bot}>0.7$&0.255 $\pm$ 0.025
&  0.232  &  0.258  &  0.260  &  0.263 &  0.246  \\
$(\Xi+\bar{\Xi})\over(\Lambda+\bar{\Lambda})$   &\footnotesize{2} &$p_{\bot}>1.$ &0.13 $\pm$ 0.03
&  0.169  &  0.114  &  0.118  &  0.122 &  0.120  \\
${K^0_{\rm s}}/\phi$          &\footnotesize{3,4} &            &11.9 $\pm$ 1.5\ \
&  6.3    &  10.4   &  9.89   &  9.69  &  16.1   \\
${K^+}/{K^-}$                                   &\footnotesize{5} &              &1.80  $\pm$ 0.10
&  1.96   &  1.75   &  1.76   &  1.73  &  1.62   \\
$p/{\bar p}$                                    &\footnotesize{6} &            &18.1 $\pm$4.\ \ \ \
&  22.0   &  17.1   &  17.3   &  17.9  &  16.7   \\
${\bar\Lambda}/{\bar p}$                        &\footnotesize{7} &              &3. $\pm$ 1.
&  3.02   &  2.91   &  2.68   &  3.45  &  0.65   \\
${K^0_{\rm s}}$/B                               &\footnotesize{3} &              &0.183 $\pm$ 0.027
&  0.305  &  0.224  &  0.194  &  0.167 &  0.242  \\
${h^-}$/B                                       &\footnotesize{3} &              &1.83 $\pm $ 0.2\ \
&  1.47   &  1.59   &  1.80   &  1.86  &  1.27   \\
${\Omega}/{\Xi}$                                &\footnotesize{1} &$p_{\bot}>0.7$&0.192 $\pm$ 0.024
&  0.119$^*$  &  0.080$^*$  &  0.078$^*$  &  0.080$^*\!\!$ &  0.192  \\
${\overline{\Omega}}/{\overline{\Xi}}$          &\footnotesize{8} &$p_{\bot}>0.7$&0.27 $\pm$ 0.06
&  0.28$^*$   &  0.17$^*$   &  0.17$^*$   &  0.18$^*$  &  0.40   \\
${\overline{\Omega}}/{\Omega}$                  &\footnotesize{1} &$p_{\bot}>0.7$&0.38 $\pm$ 0.10
&  0.55$^*$   &  0.56$^*$   &  0.57$^*$   &  0.59$^*$  &  0.51   \\
$(\Omega+\overline{\Omega})\over(\Xi+\bar{\Xi})$&\footnotesize{8} &$p_{\bot}>0.7$&0.20 $\pm$ 0.03
&  0.15$^*$   &  0.10$^*$   &  0.10$^*$   &  0.10$^*$  &  0.23   \\
\hline
& & $\chi^2_{\rm T}$ & & 88 & 24 & 1.6 & 2.7 & 19 \\
\hline\hline
\end{tabular}
\end{center}
\end{table}


The particle data shown in tables  \ref{resultsw} and \ref{resultpb} 
are obtained looking for a set of physical parameters
 which will minimize the difference between theory and 
experiment. This minimization is accomplished using the MINUIT96.03 program package 
from the CERN Fortran library. Several approaches to data analysis 
were tested in order to show the relevance of the different physical
effects we discussed above. In the approach B we keep to the
chemical equilibrium and thus allow for a search amongst three  parameters,
$T_{\rm f}$, $\lambda_{\rm q}$, and   $\lambda_{\rm s}$ leaving the two 
non-equilibrium chemical parameters at their equilibrium value ($\gamma=1$).
This is the so-called hadronic gas model, and  it fails. 
In  the approach  C we introduce strangeness chemical non-equilibrium, {\it i.e.}, we 
also vary $\gamma_{\rm s}$ and in  D we  also vary the 
light quark non-equilibrium abundance parameter  $\gamma_{\rm q}$ searching for 
best agreement between theory and experiment. In fit F we 
include the particle abundances comprising the triply strange
$\Omega,\overline\Omega$ particles.  These particles are found  in 
the S--Au/W/Au system to give a slight difficulty, despite their great measurement 
errors. In the Pb--Pb case we see  in table \ref{resultpb} 
that there is a systematic deviation 
not accounted even in the here presented non-equilibrium model. 
The particle ratios  comprising $\Omega$ and $\overline\Omega$-particles do not follow 
the same systematics. This can be also seen in
particle spectra inverse slopes presented by the WA97 collaboration \cite{WA97}.
A possible hypothesis is that  a good fraction of
these particles are made in processes that are different 
in nature than those leading to the other particle abundances.

In general we have not enforced
the strangeness conservation among the emitted hadrons in the phase space. 
Setting $\langle s-\bar s\rangle=0$ 
introduces a constraint between  parameters which is difficult 
to satisfy for S-induced reactions, where the particle abundances are 
obtained at relatively high $p_\bot$, and thus only a small fraction of all
strange particles is actually observed. This constraint is relatively easily
satisfied for the Pb--Pb collision results, where a much greater proportion 
of all strange particles is actually experimentally detected.
Fits comprising the strangeness conservation are marked by sub-script `s', and
for space reasons we only present here results for the full non-equilibrium case
D$_s$. 

We  show also at the bottom of tables~\ref{resultsw}~and~\ref{resultpb} the overall 
relative error:
\begin{equation}
\chi^2_{\rm T}=\sum_j\left(
 \frac{R_{\rm th}^j-R_{\rm exp}^j}{{\Delta R _{\rm exp}^j}}
   \right)^2\,,
\end{equation}
for all the particles considered (18 or 15 without Omegas for S--Au/W/Pb, and 
respectively 15 or 11 without Omegas for Pb--Pb). Some of the considered
data points can be obtained from others in terms of their definitions,
there are two types of relations:
\begin{equation}
\frac{\Omega+\overline{\Omega}}{\Xi+\bar{\Xi}}
=\frac{\overline\Omega}{\Xi}\cdot\,
    \frac{1+\overline\Omega/\Omega}{1+\overline\Xi/\Xi}\,,
\qquad 
\frac{\overline\Lambda}{\Lambda}=\frac{\overline\Lambda}{\overline\Xi}\cdot 
  \frac{\overline\Xi}{\Xi}\cdot \frac{\Xi}{\Lambda}\,,
\end{equation}
and for this reason  is not trivial to determine the confidence level that
goes with the different schemes B--F.  However, due to smallness of the total 
error found for the chemical nonequilibrium cases,
it is clear that only these have relatively high statistical significance,
and we do not need to go into deeper arguments  to
see which of the options presented have negligible confidence level, and which
deserve to be considered as physically interesting: for S--Au/W/Pb significant  
is case D (chemical nonequilibrium, no strangeness conservation and no
$\Omega$ in fit) and to lesser degree fit F (same, but with $\Omega$). 
Notable is that  the case D$_s$ (strangeness conservation enforced) 
fails. In   Pb--Pb  reactions again the case D is acceptable, as is the also
 D$_s$ in which we  force strangeness conservation, but case F  which includes the 
$\Omega$ data points fails. 

\begin{table}[t!]
\caption{\label{fitsw}
Statistical parameters which best describe the experimental  S--Au/W/Pb 
results  shown in table \protect\ref{resultsw}. Asterisk ($^*$)  
means a fixed (input) value or result of a constraint.
In approaches B to D, particle abundance ratios comprising $\Omega$
are not considered. 
In case D$_{\rm s}$  strangeness conservation in 
the particle  yields was enforced. 
In case F the three data-points with $\Omega$ are considered.}
\vspace{-0.2cm}\begin{center}
\begin{tabular}{|l|ccccc|c|}
\hline\hline
Fits&$T_{\rm f}$ [MeV]& $\lambda_{\rm q}$&$\lambda_{\rm s}$&
$\gamma_{\rm s}/\gamma_{\rm q}$&$\gamma_{\rm q}$& $\chi^2_{\rm T}$ \\
\hline
                    B
		     & 144 $\pm$ 2
                 & 1.53 $\pm$ 0.02
                 &   0.97 $\pm$ 0.02 
                 &   1$^*$
                 &   1$^*$
                 &   264  \\
                    C
		     &  147 $\pm$ 2 
                 & 1.49 $\pm$ 0.02
                 &   1.01 $\pm$ 0.02 
                 &   0.62 $\pm$ 0.02 
                 &   1$^*$
                 &   30  \\
                  D
		     & 143 $\pm$ 3 
                 & 1.50 $\pm$ 0.02
                 & 1.00 $\pm$ 0.02 
                 & 0.60 $\pm$ 0.02 
                 & 1.22 $\pm$ 0.06
                 & 6.5  \\
                    D$_s$
		     &  153 $\pm$ 3 
                 &   1.42 $\pm$ 0.02
                 &   1.10$^*\!\!$ $\pm$ 0.02 
                 &   0.56 $\pm$ 0.02 
                 &   1.26 $\pm$ 0.06
                 &   38  \\
                    F
		     &  144 $\pm$ 3 
                 &  1.49 $\pm$ 0.02
                 &  1.00 $\pm$ 0.02 
                 &  0.60 $\pm$ 0.02 
                 &  1.22 $\pm$ 0.06
                 &  12  \\
\hline\hline
\end{tabular}
\end{center}
\end{table}
\begin{table}[t!]
\caption{\label{fitpb}
Statistical parameters which best describe the experimental
Pb--Pb results  shown in table \protect\ref{resultpb}\,. For
definition and role of $\lambda_Q^{-1/3}$ see 
section~\protect\ref{Coulomb}. Asterisk ($^*$)  
means a fixed (input) value, or result of a constraint.
In approaches B to D, particle abundance ratios comprising $\Omega$
are not considered. 
In case D$_{\rm s}$  strangeness conservation in 
the particle  yields was enforced. 
In case F the four data-points with $\Omega$ are considered.}
\vspace{-0.2cm}\begin{center}
\begin{tabular}{|l|ccccc|c|}
\hline\hline
Fit&$T_{\rm f} [MeV]$& $\lambda_{\rm q}\lambda_Q^{1/6}$&
$\lambda_{\rm s}\lambda_Q^{1/3}$&
$\gamma_{\rm s}/\gamma_{\rm q}$&$\gamma_{\rm q}$& $\chi^2_{\rm T}$\\
\hline
                    B
                 &  142 $\pm$ 3
                 &  1.70 $\pm$ 0.03
                 &  1.10 $\pm$ 0.02
                 &  1$^*$
                 &  1$^*$
                 &  88  \\

                    C
                 & 144 $\pm$ 4
                 & 1.62 $\pm$ 0.03
                 & 1.10 $\pm$ 0.02
                 & 0.63 $\pm$ 0.04
                 & 1$^*$
                 & 24  \\
                    D
                 &    134 $\pm$ 3
                 &   1.62 $\pm$ 0.03
                 &   1.10 $\pm$ 0.02
                 &   0.69 $\pm$ 0.08
                 &   1.84 $\pm$ 0.30
                 &    1.6       \\

                    D$_s$
                 &   133 $\pm$ 3
                 &   1.63 $\pm$ 0.03
                 &   1.09$^*\!\!$ $\pm$ 0.02
                 &   0.72 $\pm$ 0.12
                 &   2.75 $\pm$ 0.35
                 &   2.7       \\

                    F
                 &    334 $\pm$ 18
                 &   1.61 $\pm$ 0.03
                 &   1.12 $\pm$ 0.02
                 &   0.50 $\pm$ 0.01
                 &   0.18 $\pm$ 0.02
                 &   19       \\
\hline\hline
\end{tabular}
\end{center}
\end{table}
The statistical parameters  yielding best particle multiplicities 
for the S-induced reactions are shown in the table \ref{fitsw}. 
In table \ref{fitpb} we show the corresponding  Pb--Pb  parameters. 
We note  that the individual phase space occupancies in Pb--Pb case
change drastically  between cases D and D$_s$, however the ratio of the 
nonequilibrium parameters
$\gamma_s/\gamma_q$ barely  changes from 0.69  to 0.72.
In actual numerical procedure we took advantage of this stability 
in $ \gamma_s/\gamma_q$-ratio. In the case of 
S--induced reactions this ratio is seen to come out slightly smaller and like
the quark fugacities, it is relatively stable for all different 
strategies to interpret the experimental results.

Notable among  the  values of statistical parameters is the remarkable result
$\lambda_{\rm s}\simeq 1$ in the S--Au/W/Pb case, see table \ref{fitsw}. 
This result requires considerable attention. We recall that the fugacities 
$\lambda_j$ arise from conservation laws, in our context, of
quark (baryon) number and strangeness in the particle source.
$\lambda_q\equiv e^{\mu_q/T}$ is thus the fugacity
of the valance light quarks. For a nucleon $\lambda_N=\lambda_q^3$,
and  hence the baryochemical potential is: $\mu_b=3\mu_q$\,.
In another refinement both $u,\,d$-flavor fugacities $\lambda_u$
and $\lambda_d$ can be introduced, allowing for up-down-quark
asymmetry. We recall that by definition $2\mu_q=\mu_d+\mu_u$,
thus $\lambda_q\equiv\sqrt{\lambda_u\lambda_d}$. We have implemented
one correction related to this refinement, since in the experimental 
data only the $\Xi^-(ssd)$ particle is measured. 

Returning again to strangeness fugacity: for strange quarks we
have $\lambda_s\equiv e^{\mu_s/T}$\,. For an antiparticle fugacity
$\lambda_{\bar i}=\lambda_i^{-1}$\,. Some papers refer in this
context to hyper-charge  fugacity $\lambda_{\rm S}=\lambda_q/\lambda_s$,
thus $\mu_{\rm S}=\mu_q-\mu_s$.
This is a highly inconvenient historical definition arising from
considerations of a hypothetical hadron gas phase. It hides from
view important symmetries, such as $\lambda_s\to 1$ for a state in which
the phase space size for strange and anti-strange quarks is the same: at
finite baryon density the number of hyperons is always greater than the
number of anti-hyperons and thus the requirement
$\langle N_s-N_{\bar s}\rangle=0$ can only be satisfied for some
nontrivial $\lambda_s(\lambda_q)\ne 1$. Thus even a small deviation from
$\lambda_s\to1$ limit must be fully understood  in order to argue that the source
is deconfined. Conversely, observation of $\lambda_s\simeq 1$ consistently
at different experimental conditions is a strong and convincing argument
that at least the strange quarks are unbound, {\it i.e.,} deconfined.

\section{Coulomb effect} \label{Coulomb}
So how do we explain the fact that Pb--Pb shows definitive deviation
from the canonical value $\lambda_s\to1$ associated with QGP? 
The answer has been keeping us for quite a while busy and only 
recently the  obvious  resolution has been found: for the highly 
Coulomb-charged fireballs formed in Pb--Pb collisions
a further effect  which needs consideration
is  the distortion of the particle phase space by the Coulomb potential. This
effect influences  particles and antiparticles in opposite way, and has
by factor two different strength for $u$-quark (charge $+2/3|e|$) and
($d,s$)-quarks (charge $-1/3|e|$).  Because Coulomb-effect acts in
opposite way on $u$ and $d$ quarks, its net impact on $\lambda_q$ is
relatively small as we shall see. 

On the other hand, the Coulomb effect 
distorts significantly the expectation regarding $\lambda_s\to 1$
for strangeness-deconfined source with vanishing net strangeness.
The difference between strange and anti-strange quark numbers
(net strangeness) allowing for a Coulomb
potential within a relativistic Thomas-Fermi phase space
occupancy model \cite{MR75}, allowing for finite temperature in QGP is:
\begin{eqnarray}\nonumber
\langle N_s-N_{\bar s}\rangle =\int_{R_{\rm f}} 
        g_s\frac{d^3rd^3p}{(2\pi)^3}&&\hspace*{-0.5cm}\left[
 \frac1{1+\gamma_s^{-1}\lambda_s^{-1}e^{(E(p)-\frac13 V(r))/T}}\right.\\
\phantom{.}\label{Nsls}\\
&-&\nonumber\left.
 \frac1{1+\gamma_s^{-1}\lambda_se^{(E(p)+\frac13 V(r))/T}}\right]\,,
\end{eqnarray}
which clearly cannot vanish for $V\ne 0$ in the limit $\lambda_s\to1$.
In Eq.\,(\ref{Nsls}) the subscript ${R_{\rm f}}$ on the spatial integral 
reminds us that only the classically
allowed region within the fireball is covered in the integration over the 
level density; $E=\sqrt{m^2+\vec p^{\,2}}$, and for a uniform charge distribution
within a radius $R_{\rm f}$ of charge $Z_{\rm f}$:
\begin{equation}
V=\left\{
\begin{array}{ll}
-\frac32 \frac{Z_{\rm f}e^2}{R_{\rm f}}
      \left[1-\frac13\left(\frac r{R_{\rm f}}\right)^2\right]\,,
          & \mbox{for}\quad r<R_{\rm f}\,;\\
 & \\
-\frac{Z_{\rm f}e^2}{r}\,,& \mbox{for} \quad r>R_{\rm f}\,.
\end{array}
\right.
\end{equation}

One obtains a rather precise result for the range of parameters of interest
to us (see below) using the Boltzmann approximation:
\begin{equation}
\langle N_s-N_{\bar s}\rangle =
\gamma_s\left\{\int g_s\frac{d^3p}{(2\pi)^3}e^{-E/T}\right\}
\int_{R_{\rm f}} d^3r\left[\lambda_s e^{\frac V{3T}}
 - \lambda_s^{-1} e^{-\frac V{3T}}\right]\,.
\end{equation}
The Boltzmann limit allows also to verify the signs: the Coulomb
potential is negative for the negatively charged $s$-quarks with
the magnitude  of the charge, $1/3$, made explicit in the potential
terms in all expressions  above. It turns out that there is always
only one solution, with resulting $\lambda_s>1$. The magnitude of the
effect is quite significant: choosing $R_{\rm f}=8$\,fm, $T=140$\,MeV,
$m_s=200$\,MeV (value of $0.5 <\gamma_s<2$ is irrelevant) solution
of Eq.\,(\ref{Nsls}) for $Z_{\rm f}=150$ yields $\lambda_s=1.10$
(precisely: 1.0983, 1.10 corresponds to $R_{\rm f}=7.87$\,fm). 
The remarkable result we found from experimental data is 
indeed this value $\lambda_s=1.10\pm0.02$, see table \ref{fitpb}. 
Thus  also in Pb--Pb system as before for the
lighter system S--Au/W/Pb \cite{Raf91,Let95} we are finding that the
source of strange hadrons  is governed by
a symmetric (up to Coulomb-asymmetry), and thus 
presumably deconfined strange quark phase space.

To better understand the situation it is convenient to introduce 
a charge conservation fugacity $\lambda_{\rm Q}$.  The abundance 
of charged quarks is thus counted by the fugacities:
\begin{eqnarray}\nonumber
\lambda_s\equiv\tilde\lambda_s\lambda_Q^{-1/3}\,, \qquad \qquad
\lambda_d\equiv\tilde\lambda_d \lambda_Q^{-1/3}\,, \qquad
\lambda_u\equiv\tilde\lambda_u \lambda_Q^{2/3}\,, \\ \label{tildl}
\lambda_q\equiv \tilde\lambda_q\lambda_Q^{1/6}=\sqrt{\lambda_u\lambda_d}\,.
\end{eqnarray}
Given that quark flavor is conserved, and thus charge is conserved 
implicitly too, $\lambda_{\rm Q}$ is not another
independent statistical variable, it is determined 
by the Coulomb potential:
\begin{equation}
\lambda_{\rm Q}\equiv 
\frac{\int_{R_{\rm f}} d^3r e^{\frac V{T}} }{\int_{R_{\rm f}} d^3r}\,.
\end{equation}

The difference between $\lambda_q$ and $\tilde\lambda_q$, see
Eq.\,(\ref{tildl}), is
at the level of 1--2\% considering the Pb--Pb case, and yet
smaller for S--A reactions.  However, the individual light quark
fugacities experience more significant shifts as we saw for $\lambda_s$.
This means that the predictions we and others have made, ignoring the 
Coulomb effect, for strange particle ratios need improvement. 
Specifically, the Pb--Pb Coulomb correction factor for the QGP-ratios is 
(previously predicted ratios $i$ are to be multiplied by $f_i$):
$f_{\overline\Xi/\overline\Lambda}=1/f_{\Xi/\Lambda}=
f_{\overline\Omega/\overline\Xi}=1/f_{\Omega/\Xi}
=\lambda_Q^{1/3}\simeq0.91\pm0.02\,,\
f_{\overline\Lambda/\Lambda}=\lambda_Q^{2/3}\simeq 0.83\pm0.03\,,\
f_{\overline\Xi/\Xi}=\lambda_Q^{4/3}\simeq 0.68\pm0.05\,,\
f_{\overline\Omega/\Omega}=\lambda_Q^{6/3}\simeq 0.56\pm0.07$\,.
This correction is required, since  previous studies of 
QGP state properties referred to the
tilde-fugacities, while data analysis where obtained for
the  non-tilde fugacities, see, {\it e.g.}, Ref.\cite{fit97}. 
Even though charge conservation has
been enforced previously \cite{BGS98}, this was done without
allowance for the Coulomb deformation of the phase space and
hence there was no net difference between tilde and non-tilde
quantities for large systems. We conclude that a QGP source 
in order to conserve strangeness has indeed $\tilde\lambda_s=1$,
which implies $\lambda_s=\lambda_{\rm Q}^{-1/3}$,  and moreover \cite{Raf91}:
\begin{equation}\label{muasym}
\frac{\langle d-\bar d\rangle}{\langle u-\bar u\rangle}=
\frac{2A-Z}{A+Z}=\frac{\tilde \mu_d}{\tilde\mu_u}
=\frac{\ln\tilde\lambda_d}{\ln\tilde\lambda_u}\,.
\end{equation}
For Pb--Pb collisions with $A=208$ and $Z=82$ there is indeed a
non-negligible isospin asymmetry, but the
Coulomb effect slightly over-compensates it and hence 
$\mu_u>\mu_d$, while $\tilde \mu_u<\tilde\mu_d$\,.

\section{Physical properties of the fireball at  chemical freeze-out} 
Given the precise statistical information about the properties of the 
hadron phase space
provided in particular by the non-equilibrium  case D, we can determine the specific 
content in energy, entropy, strangeness contained in
all hadronic particles, see tables~\ref{tqsw}~and~\ref{tqpb}.
We show here for the study cases B--F, along with their temperature the
specific energy and entropy content, and specific anti-strangeness content,
along with specific strangeness asymmetry, and finally pressure at 
freeze-out,  evaluated by 
using the best statistical parameters to characterize the hadronic particle 
phase space. We note that it is improper in general to refer to these properties 
as those of a `hadronic gas' formed in nuclear collisions, as the particles 
considered may be emitted in sequence from a deconfined source, and thus there 
never is a  stage corresponding to a hadron gas phase.

\begin{table}[tb]
\caption{\label{tqsw}
$T_{\rm f}$ and physical properties 
(specific energy, entropy, anti-strangeness, net strangeness,
pressure and volume) of  the full 
hadron phase space characterized by the statistical parameters  
given in table~\protect\ref{fitsw} for the reactions S--Au/W/Pb. 
Asterisk~$^*$ means fixed input.} 
\vspace{-0.2cm}\begin{center}
\begin{tabular}{|l|cccccc|} 
\hline\hline
Fits&$T_{\rm f}$ [MeV]& $E_{\rm f}/B$  & $S_{\rm f}/B$ & ${\bar s}_{\rm f}/B$ 
& $({\bar s}_{\rm f}-s_{\rm f})/B$ &$P_{\rm f}$ [GeV/fm$^3$] \\
\hline
                    B
		     & 144 $\pm$ 2
                 & 8.9 $\pm$ 0.5
                 &  50$\pm$ 3
                 &  1.66 $\pm$ 0.06
                 &  0.44 $\pm$ 0.02  
                 &  0.056 $\pm$ 0.005 \\
                    C
		     &  147 $\pm$ 2
                 & 9.3 $\pm$ 0.5
                 & 49$\pm$ 3
                 & 1.05 $\pm$ 0.05
                 & 0.23 $\pm$ 0.02  
                 & 0.059 $\pm$ 0.005   \\
                    D
		     &  143 $\pm$ 3
                 & 9.1 $\pm$ 0.5
                 & 48$\pm$ 3
                 & 0.91 $\pm$ 0.04
                 & 0.20 $\pm$ 0.02  
                 & 0.082 $\pm$ 0.006  \\ 
                    D$_s$
		     &   153 $\pm$ 3
                 &  8.9 $\pm$ 0.5
                 &  45 $\pm$ 3
                 &  0.76 $\pm$ 0.04
                 &  0$^*$  
                 &  0.133 $\pm$ 0.008 \\ 
                    F
		     &  144 $\pm$ 2
                 & 9.1 $\pm$ 0.5
                 & 48$\pm$ 3
                 & 0.91 $\pm$ 0.05
                 & 0.20 $\pm$ 0.02  
                 & 0.082 $\pm$ 0.006   \\
\hline\hline
\end{tabular}
\end{center}
\end{table}
\begin{table}[tb]
\caption{\label{tqpb}
$T_{\rm f}$ and physical properties 
(specific energy, entropy, anti-strangeness, net strangeness,
pressure and volume) of  the full 
hadron phase space characterized by the statistical parameters  
given in table~\protect\ref{fitpb} for the reactions Pb--Pb. 
Asterisk~$^*$ means fixed input.}
\vspace{-0.2cm}\begin{center}
\begin{tabular}{|l|lccccc|}
\hline\hline
Fit&$T_{\rm f}$ [MeV]& $E_{\rm f}/B$  & $S_{\rm f}/B$ & ${\bar s}_{\rm f}/B$ & 
$({\bar s}_{\rm f}-s_{\rm f})/B$ &$P_{\rm f}$ [GeV/fm$^3$] \\
\hline
                    B
                 & 142 $\pm$ 3
                 & 7.1 $\pm$ 0.5
                 &  41 $\pm$ 3
                 &  1.02 $\pm$ 0.05
                 &  0.21 $\pm$ 0.02
                 &  0.053 $\pm$ 0.005 \\
                    C
                 &  144 $\pm$ 4
                 & 7.7 $\pm$ 0.5
                 & 42 $\pm$ 3
                 & 0.70 $\pm$ 0.05
                 & 0.14 $\pm$ 0.02
                 & 0.053 $\pm$ 0.005 \\
                    D
                 &  134 $\pm$ 3
                 &  8.3 $\pm$ 0.5
                 &  47 $\pm$ 3
                 &  0.61 $\pm$ 0.04
                 &  0.08 $\pm$ 0.01
                 &  0.185 $\pm$ 0.012 \\
                    D$_s$
                 &   133 $\pm$ 3
                 &  8.7 $\pm$ 0.5
                 &  48 $\pm$ 3
                 &  0.51 $\pm$ 0.04
                 &  0$^*$
                 &  0.687 $\pm$ 0.030 \\
                    F
                     &  334 $\pm$ 18
                 & 9.8 $\pm$ 0.5
                 & 24 $\pm$ 2
                 & 0.78 $\pm$ 0.05
                 & 0.06 $\pm$ 0.01
                 & 1.64 $\pm$ 0.006 \\
\hline\hline
\end{tabular}
\end{center}
\end{table}


Strangeness balance is well established only in the Pb--Pb data, and thus 
it is  likely that the difficulty to balance strangeness in the 
S--Au/W/Pb is really a reflection on the high $p_\bot$ cuts inherent in these 
experimental results, which had to be obtained against the background 
of many soft spectator particles, given the non-symmetric collision system
considered.
The specific $\bar s$ content is determined to be $0.90\pm0.04$
in S--Au/W/Pb case and it is noticeably less, $0.61\pm0.04$ for the
Pb--Pb collisions. This may be perhaps attributable to greater baryon 
content of the Pb--formed fireballs, and/or, somewhat lower available 
energy, in case  D we find a difference of 0.8 GeV per nucleon in 
the CM frame. The second explanation may be more to the point
considering that in the
S--S case for which we have yet slightly higher energy available in the
collision, yet higher strangeness content can be 
qualitatively inferred \cite{SGHR94}. 
Thus we question if it could be that the energy of CERN experiments 
is just barely at the 
critical value and that every small energy increase matters? This invites
an experimental test, since very probably with sufficient effort 
more energetic Pb-SPS beams could be achieved.  
 
The specific entropy content also drops slightly for Pb--Pb ($45\pm3$)
compared with S--Au/W/Pb ($48\pm3$) for the best case D --- these values
agree well with the entropy content evaluation made 
earlier \cite{Let93}. This is so because the abundances of (predominantly high 
$p_\bot$) strange particle data are indeed found in this study to be
fully consistent within a chemical-nonequilibrium description  with the $4\pi$
total particle multiplicity results. The pressure of the hadronic phase space
in high confidence cases $P_{\rm f}\simeq 0.1$--$0.2\,{\rm GeV/fm}^3$
has the magnitude of the vacuum  confinement pressure (bag constant).

We do not present here the highly strategy sensitive freeze-out
volume $V_{\rm f}$, which can be determined assuming that the hadronic
phase space comprises for the S--Au/W/Pb collisions baryon number $B\simeq 120$,
and in Pb--Pb case being $B \simeq 372\pm10 $, 
as stated in \cite{BGS98}. However, we find that a spherical source 
corresponding to the best case D in case of Pb--Pb reactions
would have a source radius 9.6 fm, which in turn can be
checked to be in agreement with deconfined strangeness conservation
as described by Eq.\,(\ref{Nsls}), given the established statistical parameters 
and $m_s=200$\,MeV.

\begin{figure}[tb]
\vspace*{0.3cm}
\centerline{\hspace*{4.8cm}
\psfig{width=11.2cm,figure=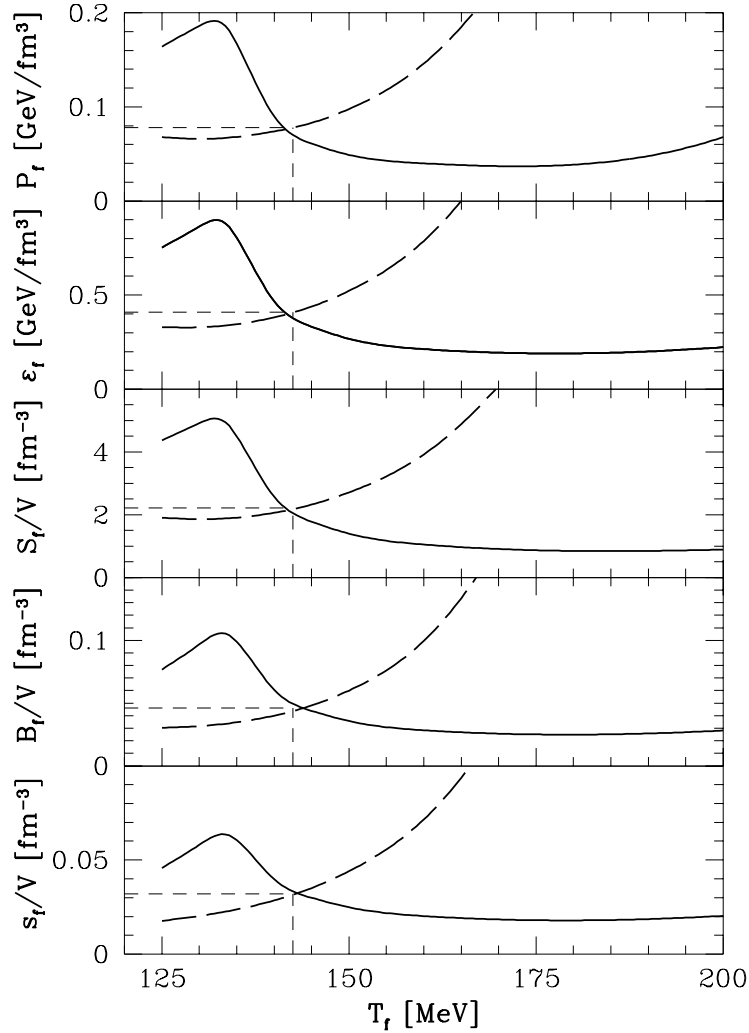}
}
\vspace*{-2.cm}
\caption{Comparison of the physical properties of the hadron source 
in  Pb--Pb (solid lines) and S--Au/W/Pb (dashed lines) collision fireballs.
Curves are obtained minimizing disagreement between theory and 
the experimental particle  abundance data for a given $T_{\rm f}$.
We show in sequence from top to bottom: pressure $P_{\rm f}$, 
energy density $\epsilon_{\rm f}$ entropy density, $S_{\rm f}/V$, 
baryon density $B_{\rm f}/V$ and $s$-quark density $s_{\rm f}/V$.
\protect\label{compare}
}
\end{figure}

Comparing the two tables~\ref{tqsw}~and~\ref{tqpb} we note 
similarities in results which become more clear when we study 
the  extensive physical properties of the two freeze-out 
systems. We choose a fixed given
value of the freeze-out temperature $T_{\rm f}\in$ (125--200) MeV and 
find best values of the  other parameters according to the strategy  D, 
and evaluate for both systems 
extensive physical properties.  This comparison yields 
a surprise shown in figure~\ref{compare}. We see that
solid lines (case of Pb--Pb) are crossing dashed lines (case of S--Au/W/Pb) at 
nearly the same value of $T_{\rm f}\simeq 143$ MeV. Thus it appears
possible that a universal freeze-out for these and other similar collision 
systems arises,  corresponding possibly to the common
physical properties of QGP at its breakup into hadrons --- in this context we draw
attention to the interesting result about the scale of  the 
energy density at freeze-out: $\epsilon_{\rm f}=0.43\pm 0.04 \,{\rm GeV/fm}^3$. 
These results, along with earlier shown strange phase space symmetry and
the Coulomb effect in the Pb--Pb system
 has as simple interpretation the formation of a
deconfined phase in the initial stages of the collision, which subsequently
evolves and flows apart till it reaches the universal hadronization point, with 
many similar physical properties, independent of the collision system. 
System dependent will certainly the surface collective 
velocity~$\vec v_{\rm c}$. 

\section{Collective flow}\label{flowsec}

So far we have considered only  the particle yields, and have
not addressed particle spectra which 
requires that we allow for the Doppler-like
blue shift of the particles emitted from a moving source at hadronization.
To do this quantitatively we need to consider explicitly
 the collective  velocity $\vec v_{\rm c}$, 
see section \ref{subsecflow}. 
Moreover, while the integral over  the entire phase space of the flow spectrum 
yields as many particles with and without flow, when acceptance cuts are 
present particles of different mass experience differing flow effects. 
While we made an effort to avoid forming ratios of particles which are subject
to greatly differing flow effects, we must expect some change in our results.
We will consider here the radial flow model, perhaps of the  simplest of
the complex flow cases possible \cite{Hei92}, 
but it suffices to fully assess the impact of flow on our analysis.

We deal with the Doppler effect as follows:
for a given pair of values $T_{\rm f}$ and $v_{\rm c}$, the resulting  
$m_\bot$ particle spectrum is obtained and analyzed using the spectral shape 
and procedure (error is proportional to spectral strength) 
employed by the experimental groups, and 
the theoretical inverse slope `temperature' $T_{\rm s}^j$ is determined 
for each particle $j$  which we can compare with the experimental 
results for $T_{\rm s}^j$\,. 
Let us first address the case of S--W reactions. Here the strange (anti)baryon
slopes are quite similar \cite{Eva96}. In consideration of, within error, 
overlapping values of $T_{\rm s}^j$ we decided to consider only one  value
$T_{\rm s}=235\pm10$, chosen near to the most 
precise fitted Lambda spectra slope. Once we find values of $T_{\rm f}$
and $v_{\rm c}$, we  can check how the slopes of all 
 particles have fared. The resulting 
$T_{\rm s}^j$ are  in remarkable agreement with experiment, well beyond
what we expected: we find for kaons, lambdas  and cascades the 
values $T_{\rm s}^j=215, 236$ and 246 MeV respectively,
which both in trend and value agrees with the K$^0$, 
$\Lambda,\,\overline\Lambda$, $\Xi$ and $\overline\Xi$ WA85 
results~\cite{Eva96}: $T_{\rm s}^{{\rm K}^0}=219\pm5,\ 
T_{\rm s}^\Lambda=233\pm3,\
T_{\rm s}^{\overline\Lambda}=232\pm7\,,\ T_{\rm s}^\Xi=244\pm12$ and 
$T_{\rm s}^{\overline\Xi}=238\pm16$.  

Since the flow effect shifts particles of different mass differently
into different domains of $m_\bot,\,y$\,, it is not surprising that the
inclusion of flow only impacts the phase space abundance parameters,
beyond their established  errors. In the theory -- experiment comparison the 
strategy corresponding to the case D remains the best, it has now
(eliminating to simplify  the analysis  all redundant data points)
$\chi^2/$dof = 0.73.  Fit F has $\chi^2/$dof = 0.83, and thus we can
conclude that the S--Au/W/Pb data is fully understood within the freeze-out
non-equilibrium model with flow.  In order to facilitate 
comparison with other work we note here  to greater mathematical 
precision our  case D with flow results for S--Au/W/Pb: 
$T=142.7\pm2.1$ MeV, $\mu_b=176\pm3$ and 
$|\vec v_{\rm c}|=0.486\pm0.010\,c$. We also note that 
allowing for flow the consequences regarding strangeness
non-conservation remain unchanged: case D$_s$ with flow has 
negligible confidence level with $\chi^2$/dof $\simeq 3.2$. The 
strangeness imbalance in fits D, F is the same as we have 
obtained without flow. We conclude that strangeness imbalance problem 
we see in the data is not result of radial flow effects. 

An interesting feature of case D with flow is that there 
is little  correlation between now 6 theoretical parameters, in other
words the flow velocity is a truly new degree of freedom required by
the experimental data and it helps attain a good agreement with experimental 
results. We also  checked that  nearly the same flow 
velocity is found when we only study the particle multiplicity, and 
disregard the information about the experimental 
inverse slope of $m_\bot$-spectra. It is for this reason that we  present here the
simple radial flow model, as within this scheme the inverse transverse
slope of hadrons is correctly `predicted' by the chemical freeze-out 
analysis with flow, and the systematic change of the individual 
Doppler-shifted slopes seems to be as described above, just right for 
different particles. 

We note that in just one physical aspect consideration of collective 
 flow offers a new insight:  the value of $\gamma_s$ we determine
is compatible with  unity. This value was noted already in the 
analysis of the S--S collision results \cite{SGHR94}, and so far eluded 
the analysis of S--W/Au/Pb  collisions.  Specifically,
we find  $\gamma_{\rm s}/\gamma_{\rm q}=0.69\pm0.03$\,,
$\gamma_{\rm q}=1.41\pm0.08$. Another  result to note is that the 
specific energy per baryon when flow is present just 
needs the Lorentz factor $\gamma_v$ 
to be nearly exactly the same as the result we found without flow, which 
was in fact just the available energy in the collision. This observation 
merely proves time and again the physical consistency of the statistical
approach to particle abundance study. 

We are presently completing a similar analysis of the Pb--Pb system which 
shall be reported in another paper \cite{LRflow98}. 
Our findings are for this reaction case also fully compatible
with the picture of the reaction we have been developing here, 
the major highlights of this work are that the collective 
radial flow velocity is fitted now to be just above, but compatible with
the sound velocity of quark matter 
$v_{\rm c}\simeq c/\sqrt{3}\simeq 0.6\,c$. The chemical non-equilibrium
parameters $\gamma_{q,s}$ are yet bigger, and exceed both value 2,
the temperature at freeze-out increases slightly to 136--139 MeV from the 
value 133 MeV we saw in the case D without flow. The
fugacities $\lambda_{q,s}$ remain unchanged. Despite further increase
in the values of $\gamma_{q,s}$  
we still find a smaller specific yield of $\bar s/b\simeq 0.7$ compared
to the S--Au/W/Pb case. The higher  collective velocity suggests as
possible explanation the fact that the greater internal pressure reached in 
Pb--Pb collisions compared to S--Au/W/Pb case shortens the effective available
time for strangeness production, resulting in lesser yield, despite a
expected higher initial temperature \cite{acta96}. A full report on these
developments will also contain a possible resolution
of $\Omega,\,\overline\Omega$ high yield riddle exploiting a model involving 
gluon assisted production mechanism \cite{LRflow98}.

\section{Current status and conclusions}\label{secsum}
We have presented detailed analysis of hadron abundances observed
in central S--Au/W/Pb 200 A GeV and Pb--Pb 158 A GeV interactions
within thermal equilibrium and chemical non-equilibrium  phase 
space model of strange and non-strange hadronic particles.
We assumed formation of a thermal dense matter fireball of a priori
unknown structure, 
which explodes and disintegrates into the final state hadrons. 
This approach  allows excellent description of all abundance
data, and when flow is considered, also a good understanding of
 the transverse mass inverse slopes. For Pb--Pb system we have, 
depending on strategy we adopt, no less  than  5 independent
 degrees of freedom, and a few more in  S--Au/W/Pb reactions, and thus 
our approach is not a process of fitting an `elephant' to a few data, 
but indeed it should be seen as a solid confirmation 
the chemical freeze-out as a well defined stage of the evolution
of dense matter. We also find results that may indeed be of quite considerable
importance for the search for quark-gluon plasma. In particular,
the physical statistical parameters
obtained here characterize a strange particle source which, both for 
S--Au/W/Pb and for  Pb--Pb case, when allowing for Coulomb deformation 
of the strange  and anti-strange quarks,
is exactly symmetric between $s$ and $\bar s$ quark carriers, 
as is natural for a deconfined state.

We find  a highly significant description of all experimental
data leading to a fireball having specific baryon energy $E/B\simeq 9$ GeV,
high specific entropy $S/B\simeq 45$--$50$, and chemical
freeze-out temperature $T_{\rm f}\simeq 140$ MeV. The dense  blob of 
matter was  expanding in case  of S--Au/W/Pb reactions with surface 
velocity $v_{\rm c}\simeq 0.49\,c$ and in case of Pb--Pb
reactions just nearly with sound velocity of quark matter 
$v_{\rm c}\simeq c/\sqrt{3}\simeq 0.6\,c$. The near equilibrium abundance of 
strange quarks ($\gamma_{\rm s}\simeq 1$, including flow), and the
over-abundance of light quarks ($\gamma_{\rm q}^2\simeq 2$), is 
pointing to a deconfined, fragmenting quark-gluon fireball as the
direct particle emission source.

 Even though
there is still considerable uncertainty about other freeze-out 
flow effects, such as longitudinal
flow (memory of the collision axis), the level of consistency
and quality of agreement between a wide range of experimental data
and our chemical non-equilibrium, thermal equilibrium statistical 
model suggests that, for the observables considered 
here, these effects do not matter. Considering the quality of 
the data description obtained it is impossible to consider the 
 results presented here
as accidental and likely to see further major revision. 

In conclusion, we have shown that 
strange particle production data, combined with the global 
hadron multiplicity (entropy), can be consistently 
interpreted within a picture of a hot hadronizing blob of matter
governed by statistical parameters acquiring values 
expected if the source structure
is that of deconfined QGP. We have further found that radial flow allows
to account exactly for the difference between freeze-out temperature 
and the observed spectral shape, and allows full description
of the inverse slope of $m_\bot$ strange baryon and antibaryon spectra
for S-induced reactions. 
The only natural interpretation of our findings is that these
particles are  emerging directly from hadronizing deconfined state
and do not undergo a chemical re-equilibration after they have been 
produced.

{\vspace{0.5cm}\noindent\it Acknowledgments:\\}
We thank E. Quercigh for interesting and stimulating discussions.
This work was supported in part
by a grant from the U.S. Department of
Energy,  DE-FG03-95ER40937\,. LPTHE, Univ.\,Paris 6 et 7 is:
Unit\'e mixte de Recherche du CNRS, UMR7589.



\section*{References}
\begin{thebibliography}{99}\small
\setlength{\itemsep}{-.01cm}

\bibitem{LR98}
J. Letessier and J. Rafelski, submitted to {\it Phys. Rev. C};
[hep-ph/9806386], June 16, 1998\,.

\bibitem{fit97}
J. Letessier, J. Rafelski and A. Tounsi,
 {\it Phys. Lett.} {\bf B 410}, (1997) 315; [hep-ph/9710310];\\
\hspace*{-0.38cm}J. Rafelski, J. Letessier, and A. Tounsi,
{\it Acta Phys. Polon.}, B{\bf 28}, 2841, (1997); [hep-ph/9710340].

\bibitem{LRPb98}
J. Letessier and J. Rafelski, [hep-ph/9807346], July 11, 1998\,.

\bibitem{HAG}
R. Hagedorn, Suppl. Nuovo Cimento
{\bf 2}, 147 (1965);
 Carg\`ese lectures in Physics,
Vol.\,6, Gordon and Breach (New York 1977)
and references therein. \\
See also: J. Letessier,
H. Gutbrod and J. Rafelski, {\it Hot Hadronic Matter},
NATO-ASI series B34,6  Plenum Press, New York 1995.

\bibitem{HAGBB} H. Grote, R. Hagedorn and J. Ranft,
{\it Atlas of Particle Production Spectra},
(CERN-Service d'Information Scientifique, Geneva 1970).

\bibitem{acta96}
{J. Rafelski, J. Letessier and A. Tounsi},
{\it Acta Phys. Pol.} B {\bf 27}, 1035 (1996), and references therein.

\bibitem{BSWX96}
P. Braun-Munzinger, J. Stachel, J.P. Wessels and N. Xu,
{\it Phys. Lett.} B {\bf 365}, 1 (1996); [nucl-th/9508020].

\bibitem{Eva96} 
D. Evans for the WA85 Collaboration, APH N.S., Heavy Ion Physics {\bf 4},
79 (1996) [proceedings of Strangeness 1996 -- Budapest meeting].

\bibitem{WA97}
E. Andersen {\it et al.}, WA97-collaboration, 
preprint CERN-EP-98-064, Apr 1998;  {\it Phys.Lett.} B. in press;\\ 
\hspace*{-0.3cm}R. Caliandro, WA97 collaboration, in this volume. 

\bibitem{Bas98}
S.A. Bass {\it et al.}, {\it Prog. Part. Nucl. Phys.} {\bf 41}, 225
 (1998); [nucl-th/9803035].


\bibitem{Raf91}
{J. Rafelski}, {\it Phys. Lett. }B {\bf 262}, 333 (1991);
{\it Nucl. Phys.} A {\bf 544}, 279c (1992).

\bibitem{Let95}
J. Letessier, A. Tounsi, U. Heinz, J. Sollfrank and J. Rafelski,
{\it Phys.\ Rev.} D {\bf 51}, 3408 (1995), and references therein;
[hep-ph/9212210].

\bibitem{Sol97}
J. Sollfrank, {\it J. Phys. } G {\bf 23}, 1903 (1997); 
[nucl-th/9707020], and references therein.

\bibitem{BGS98}
F. Becattini, M. Gazdzicki and J. Sollfrank,
{\it Eur. Phys. J.} {C} {\bf 5}, 143 (1998); [hep-ph/9710529].

\bibitem{Let93}
J. Rafelski, J. Letessier and A. Tounsi,
Dallas--ICHEP (1992) p.\,983 (QCD161:H51:1992); [hep-ph/9711350];\\
\hspace*{-0.3cm}J. Letessier, A. Tounsi, U. Heinz, J. Sollfrank and J. Rafelski
{\it Phys. Rev. Lett.} {\bf 70}, 3530 (1993); [hep-ph/9711349].

\bibitem{entro}
{J. Letessier, A. Tounsi and J. Rafelski},
{\it Phys. Rev. }C {\bf 50}, 406 (1994); [hep-ph/9711346];\\
\hspace*{-0.3cm}{J. Rafelski, J. Letessier and A. Tounsi},
{\it  Acta Phys. Pol.} A {\bf 85}, 699 (1994).

\bibitem{RM82}
{J. Rafelski and B. M\"uller}, {\it Phys. Rev. Lett}
{\bf 48}, 1066 (1982); {\bf 56}, 2334E (1986).

\bibitem{BZ82}
 T.S. Biro and  J. Zimanyi
{\it Phys. Lett.} B {\bf 113}, 6 (1982);
{\it Nucl. Phys.} A {\bf 395},525 (1983).

\bibitem{Hei92} 
E. Schnedermann, J. Sollfrank and 
U. Heinz, pp175--206 in {\it Particle Production in Highly Excited 
Matter}, NATO-ASI Series B303,  H.H. Gutbrod and J. Rafelski, Eds., 
(Plenum, New York, 1993)

\bibitem{MR75}
B. Muller and J. Rafelski,
{\it Phys. Rev. Lett.} 34, 349 (1975).

\bibitem{SGHR94}
J. Sollfrank, M. Ga\'zdzicki, U. Heinz and J. Rafelski,
{\it Z. Phys.} C {\bf 61}, 659 (1994).

\bibitem{LRflow98}
J. Letessier and J. Rafelski, 
``Study of collective transverse flow and of gluon assisted 
$\Omega/\overline\Omega$ production in Pb--Pb  collisions'', 
in preparation.


\end{thebibliography}
\end{document}